\documentclass[sigconf]{acmart}
\usepackage{listings}
\usepackage{multirow}

\AtBeginDocument{%
  \providecommand\BibTeX{{%
    \normalfont B\kern-0.5em{\scshape i\kern-0.25em b}\kern-0.8em\TeX}}}

\copyrightyear{2021}
\acmYear{2021}
\setcopyright{ccby}
\acmConference[K-CAP '21] {Proceedings of the 11th Knowledge Capture Conference}{December 2--3, 2021}{Virtual Event, USA.}
\acmBooktitle{Proceedings of the 11th Knowledge Capture Conference (K-CAP '21), December 2--3, 2021, Virtual Event, USA}
\acmPrice{}
\acmISBN{978-1-4503-8457-5/21/12}
\acmDOI{10.1145/3460210.3493546}

\begin{document}
\fancyhead{}

\title{User-friendly Composition of FAIR Workflows in a Notebook Environment}


\author{Robin A Richardson}
\orcid{0000-0002-9984-2720}
\affiliation{%
  \institution{Netherlands eScience Center}
  \city{Amsterdam}
  \country{Netherlands}
}

\author{Remzi Celebi}
\orcid{0000-0001-7769-4272}
\affiliation{%
  \institution{Maastricht University}
  \city{Maastricht}
  \country{Netherlands}
}

\author{Sven van der Burg}
\orcid{0000-0003-1250-6968}
\affiliation{%
  \institution{Netherlands eScience Center}
  \city{Amsterdam}
  \country{Netherlands}
}

\author{Djura Smits}
\orcid{0000-0003-4096-0260}
\affiliation{%
  \institution{Netherlands eScience Center}
  \city{Amsterdam}
  \country{Netherlands}
}

\author{Lars Ridder}
\orcid{0000-0002-7635-9533}
\affiliation{%
  \institution{Netherlands eScience Center}
  \city{Amsterdam}
  \country{Netherlands}
}

\author{Michel Dumontier}
\orcid{0000-0003-4727-9435}
\affiliation{%
  \institution{Maastricht University}
  \city{Maastricht}
  \country{Netherlands}
}

\author{Tobias Kuhn}
\email{t.kuhn@vu.nl}
\orcid{0000-0002-1267-0234}
\affiliation{%
  \institution{Vrije Universiteit Amsterdam}
  \city{Amsterdam}
  \country{Netherlands}
}

\renewcommand{\shortauthors}{Richardson et al.}

\begin{abstract}
There has been a large focus in recent years on making assets in scientific research findable, accessible, interoperable and reusable, collectively known as the FAIR principles. A particular area of focus lies in applying these principles to scientific computational workflows. Jupyter notebooks are a very popular medium by which to program and communicate computational scientific analyses. However, they present unique challenges when it comes to reuse of only particular steps of an analysis without disrupting the usual flow and benefits of the notebook approach, making it difficult to fully comply with the FAIR principles. Here we present an approach and toolset for adding the power of semantic technologies to Python-encoded scientific workflows in a simple, automated and minimally intrusive manner. The semantic descriptions are published as a series of nanopublications that can be searched and used in other notebooks by means of a Jupyter Lab plugin. We describe the implementation of the proposed approach and toolset, and provide the results of a user study with 15 participants, designed around image processing workflows, to evaluate the usability of the system and its perceived effect on FAIRness. Our results show that our approach is feasible and perceived as user-friendly. Our system received an overall score of 78.75 on the System Usability Scale, which is above the average score reported in the literature.
\end{abstract}



\maketitle

\section{Introduction}
The reproducibility of research, which is a main principle in open science, depends on the availability of original data, details about the applied methods, and how they connect to the primary results. The FAIR principles describe a set of guidelines to increase sharing and publication of research data by making it Findable, Accessible, Interoperable, and Reusable~\cite{wilkinson2016fair}. The FAIR principles have been well established for data sets that are input or output of research processes; however, it is less well determined what these principles mean for the scientific workflows themselves, although significant attempts are in progress with regards to FAIR software~\cite{lamprecht2020towards}. Sharing of workflows in more detail, including information about the individual steps, their parameters, and provenance, is essential for reproducibility. In our previous work \cite{celebi2020towards}, we proposed a semantic data model that can describe the workflow steps and their execution order (prospective provenance) and the concrete activities that happened during execution (retrospective provenance). Building upon this earlier work, we present here a concrete implementation and user interface, and show evaluation results that demonstrate the feasibility of this approach.

Jupyter notebooks \cite{kluyver2016jupyter} have become popular in the data science community due to their flexibility and ability to create and test data science workflows and view results interactively. There are a number of efforts to address the sharing of workflows using workflow languages (e.g., CWL, WDL and GWL) or workflow management systems (e.g., Galaxy \cite{goecks2010galaxy}, KNIME \cite{berthold2009knime} and Nextflow \cite{di2017nextflow}) \cite{strozzi2019scalable}. Most existing workflow languages and approaches deliberately separate the individual step implementations from the topological definition. On the other hand, approaches such as noWorkflow \cite{murta2014noworkflow} exist, which help to capture provenance of normal Python scripts, using abstract syntax tree analysis to determine the `workflow' that has been executed (and has also been applied directly to IPython notebooks \cite{pimentel2015collecting}). The former approach tends to interfere with the `notebook way' of doing data science, whereas the latter provides less scope for reuse of parts of the workflow. We therefore hypothesize that a structured approach allowing for communicable and reusable steps, but without upsetting the natural usage of notebooks, would be beneficial in terms of interoperability as well as usability.

To make computational workflows and their individual steps truly interoperable and reusable across systems, we need to represent their nature and relations with formal semantics. This in turn allows for smart tools to help us with workflow discovery and composition, with validating workflows, and with analyzing workflows and their steps as reusable entities.

Semantic annotation of workflows can be a time-consuming and tedious process, but many of these annotations can be automatically inferred from the code itself. For these reasons, we aim to weave workflow FAIRification into the normal `flow' of workflow creation and execution by a typical data scientist, automating semantic annotation as much as possible. In this way, we seek to address a core research question: \textit{Can FAIR computational workflows be created in a notebook environment in a user-friendly manner with minimal adjustments from normal notebook usage?}

Our contributions are the following: 1) We present a framework consisting of Python libraries and Jupyter extensions that allows for easy transformation of Python functions into FAIR workflow steps by generating semantic and metadata information. 2) We apply the existing nanopublication technology \cite{groth2010anatomy,kuhn2016decentralized,kuhn2021semantic} and reuse existing semantic data models for the FAIR representation and publication of semantic entities of workflows so that the published workflow steps can be searched and loaded into the notebook. 3) We report a usability study which shows that the framework is user-friendly and suitable for composing real-life workflows such as image processing workflows.

A more thorough explanation of the relevant terms and technologies is given in section \ref{sec:background}, while the general approach behind the FAIR toolset is detailed in section \ref{sec:approach}. The design and results of a user evaluation are provided in sections \ref{sec:userstudydesign} and \ref{sec:results}. Finally, discussion of the limitations of the study and future development of the tools are given in section \ref{sec:discussion}.

\section{Background}\label{sec:background}

\subsection{Notebooks}
Notebooks, in the sense used in Data Science, are interactive programming environments that mix executable code with rich text explanations and results, including plots, images and other visual media, generally presented as a linear sequence of (content) cells. They provide an important and popular means of Literate Programming, particularly in the Data Science community. Jupyter Notebooks \cite{kluyver2016jupyter} (formerly more restrictively known as IPython notebooks) are a highly popular tool with support for many languages common in data science such as R and Python. Due to their natural application as literate programming documents, they have shown promise as a communication medium for results that can alleviate some current problems with reproducibility in science~\cite{rule2019ten}, subject to certain best practices, for example as derived from studies of millions of notebooks\cite{pimentel2019large}.

\subsection{Scientific Workflows}
In general, a scientific workflow can be defined as: "{...} the description of a process for accomplishing a scientific objective, usually expressed in terms of tasks and their dependencies." \cite{ludascher2009}. With regard to computational scientific workflows, particularly in the Bioinformatics domain, a distinction is generally drawn between workflows and software programs, considering workflows as a form of `actionable data' providing a high level description, separate from the specifics of execution which may, in principle, have more than one low level implementation. Further categorisation of workflows by their level of abstraction and role in the scientific life cycle are current active discussion topics \cite{lamprecht2021perspectives}. In practice, the line between workflow definition and implementation is not always so clear, particularly with respect to Python scripting and notebooks, which tend to blur the definition, implementation and execution concerns. Nevertheless, in this work we seek a means to structure and FAIRify such "computational workflows", therefore adopting the term in a somewhat broader sense than that commonly used in the context of workflow management systems.

An enormous number of technologies exist for the description, execution and management of computational workflows, including popular workflow management systems like Galaxy \cite{goecks2010galaxy} and Knime \cite{berthold2009knime}, and languages (one example being the Common Workflow Language CWL \cite{amstutz2016common}).


\subsection{FAIR Workflows}
`FAIR' refers to a series of guiding principles for making scientific data (and software) Findable, Accessible, Interoperable and Reusable \cite{wilkinson2016fair}. It is seen as key to improving the scientific process, for example with respect to the greatly discussed `Reproducibility Crisis' in science \cite{baker20161}.

There have been significant recent efforts to FAIRify various aspects of scientific computational workflows. In practice, the very large number of workflow technologies and standards has made this a non-trivial task \cite{ferreira_2021}.

The recent development of Research Object Crates (RO-Crate) \cite{carragain2019lightweight}, a standard for packaging research data with its metadata, has also been adopted for the storage of computational workflows to aid their findability and reusability, notably as part of the WorkflowHub \cite{workflowhub}, a workflow registry developed by the European Open Science Cloud for Life Sciences. It has also been recognised that the burden of generating large amounts of metadata is likely to hinder the FAIRification process, and therefore tools that facilitate automatic annotation will be essential \cite{goble2020fair}.

Semantic Web Technologies are a set of methods and tools that arose out of the Semantic Web field, most notably Linked Data and the associated W3C standard RDF \cite{hitzler2021review}. These technologies are key in FAIRification of data and scientific workflows, as they enable machine-readable and interoperable metadata. For some time, researchers have been combining existing ontologies to describe workflows and help prevent their `decay' into non-reusability\cite{belhajjame2015using}. In our previous work \cite{celebi2020towards}, we proposed a semantic data model that can describe the workflow steps and their execution order (prospective provenance) and the concrete activities that happened during execution (retrospective provenance). Here we continue on from the model we developed, to facilitate the generation and publication of such semantic descriptions.

\subsection{Nanopublications}
Nanopublications \cite{groth2010anatomy,kuhn2013broadening} are a formalized and machine-readable way of communicating the smallest possible units of publishable information. This could be, for example, the outcome of a scientific study or a claim made by a particular scientist. Nanopublications are searchable, citable, and contain authorship and attribution information. Nanopublications are thus a convenient and FAIR way to package up RDF, thereby making them the ideal candidate for publication of metadata and semantic annotations of scientific workflows as well individual workflow steps.

\section{Approach and Implementation}\label{sec:approach}
\begin{figure}
  \includegraphics[width=0.7\linewidth]{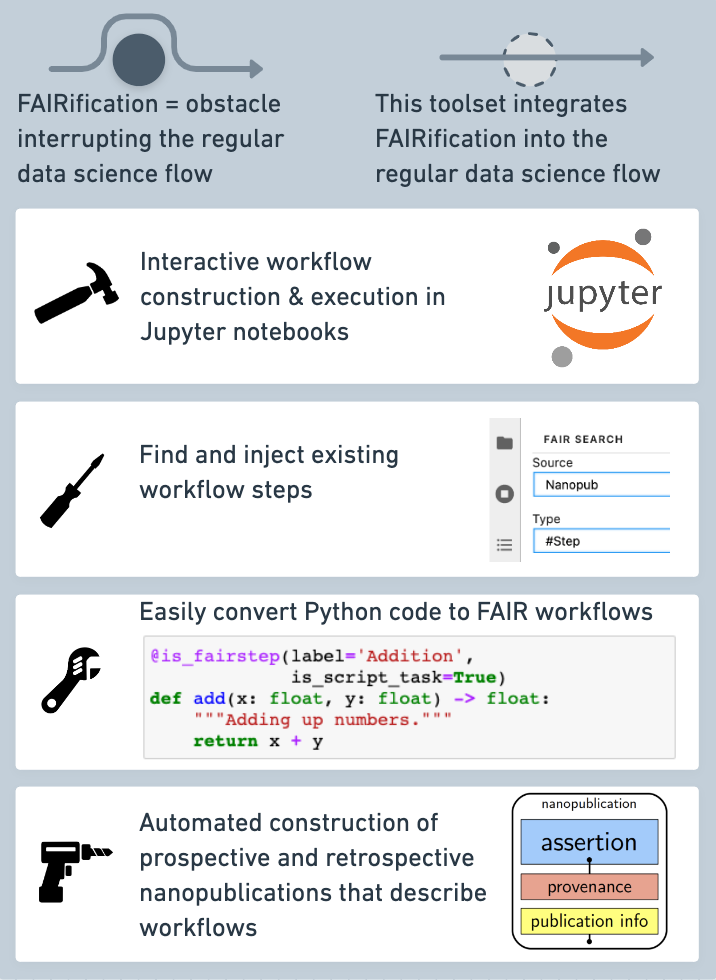}
  \caption{Overview of the set of tools presented in this work that together add the power of semantic technologies to Python encoded scientific workflows in a simple, automated and minimally intrusive manner.}
  \label{fig:approach}
\end{figure}
Figure \ref{fig:approach} gives a high level overview of the key elements of our approach. The essence of the problem is that programming in Jupyter Notebooks is attractive precisely because of its simplicity and linearity. Rearrangement of code blocks, workflow boilerplate code and manual semantic annotations are not natural to this environment and defeat the purpose of using it. We aim to add automated semantic annotation, workflow construction and publishing in as non-intrusive a manner as possible. We also aim to add a graphical interface to the notebook environment that allows searching of previously published workflow information.

To this end, we designed a set of tools to allow this creation of FAIR scientific workflows within a Jupyter notebook environment. The components of this toolset consist of:
\begin{itemize}
  \item The \texttt{nanopub} library provides a Python client for searching, publishing and modifying nanopublications. 
  \item The \texttt{fairworkflows} library supports the construction, execution and publishing of FAIR scientific workflows using semantic technologies.
  \item The \texttt{FAIRWorkflowsExtension}, a Jupyter-Lab extension to aid in the use of the \texttt{fairworkflows} library in notebooks. Allows searching and loading of existing (nano-)published steps and workflows, as well as publishing.
\end{itemize}

\begin{figure}
  \includegraphics[width=0.8\linewidth]{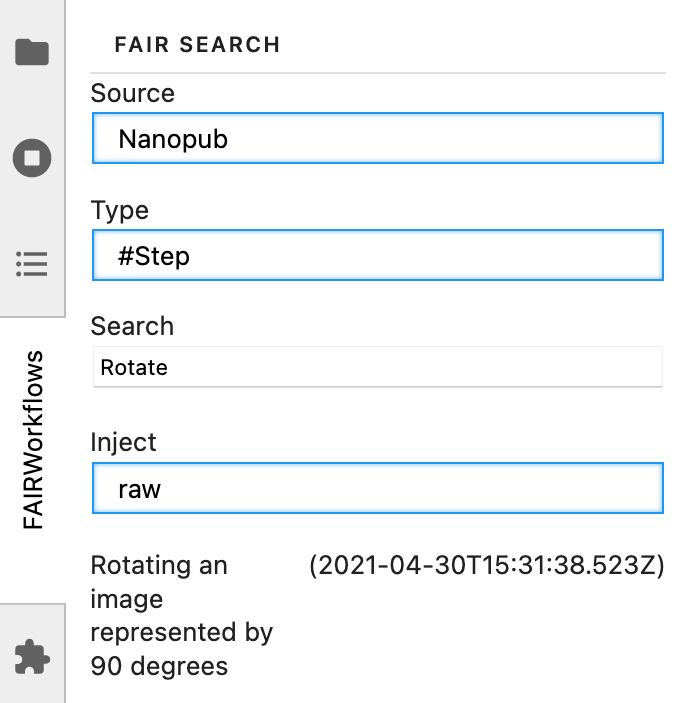}
  \caption{The Jupyter Lab plugin provides a widget with a search bar returning results from the nanopub servers.}
  \label{fig:fairworkflows_widget_search}
\end{figure}

Typical usage is as follows: 
A user would describe computational steps in a Jupyter notebook using Python functions, or search and load existing steps using the JupyterLab extension \cite{fairworkflowsextension}, whose frontend component is shown in Figure~\ref{fig:fairworkflows_widget_search}.
These `step' functions are called from one `main' function, thereby defining the workflow.
Metadata and semantic annotations are added automatically or configured at the code-level.
The workflow and steps are published from within the notebook environment.

\subsection{Nanopub Python library}
In the approach taken in the present work, annotated workflows are published as nanopublications. To facilitate searching, publication, and modification of nanopublications we developed a high-level, user-friendly Python interface: \texttt{nanopub}~\cite{nanopub2021}. In this work users only indirectly use the \texttt{nanopub} library, since it is used in the backend of the \texttt{fairworkflows} library.

Latest versions of \texttt{nanopub} can be found on the Python Package Index (\texttt{pip install nanopub}) and development on GitHub~\cite{nanopub-github-2021}.

\begin{figure}
  \includegraphics[width=\linewidth]{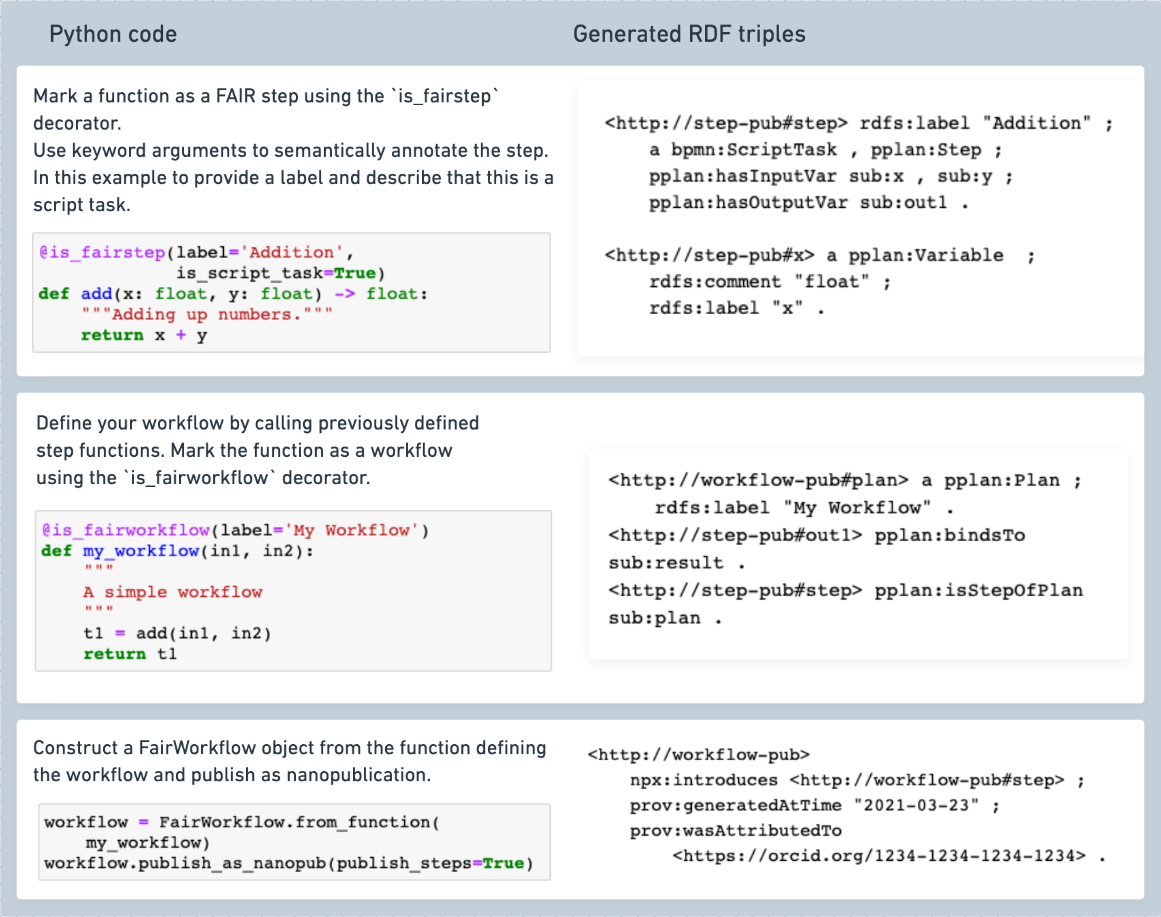}
  \caption{Diagram illustrating the usage of the Python \texttt{fairworkflows} library (on the left) and the corresponding generated RDF triples (on the right). For readability reasons we only present relevant RDF triples and leave out namespaces.}
  \label{fig:python_vs_rdf}
\end{figure}

\subsection{FAIR Workflows library}
The \texttt{fairworkflows} library~\cite{fairworkflows-library-2021} provides helpful functions and classes to convert Python code into rich semantically annotated, executable workflows. The \texttt{is\_fairstep} decorator is used to mark a normal Python function as a computational “step” whose action is defined by the function code. A function that calls multiple such decorated step functions can be decorated with the \texttt{is\_fairworkflow} decorator to mark it as a Plex workflow. Both decorators accept intuitive key-value pairs to provide further semantic annotation of the concerned steps and workflow.

Under the hood, these decorators construct RDF graphs that semantically describe the steps and workflow using the appropriate ontologies, as shown in Figure \ref{fig:python_vs_rdf}. This RDF is contained within \texttt{FairStep} and \texttt{FairWorkflow} Python objects, which provide an intuitive API to information stored in the RDF. The \texttt{FairStep.inputs} attribute, for example, returns the inputs for a FAIR step.

The resulting workflow can be published by calling the function \texttt{publish\_as\_nanopub()}, which makes use of the \texttt{nanopub} Python library presented above to generate Nanopublications containing step or workflow RDF in the assertion, publishing them to the international nanopub server network.

Additionally, the workflow can be executed by calling its method \texttt{execute()} with the desired input parameters. This returns the result of the execution (if any), as well as a \texttt{WorkflowRetroProv} object that contains an RDF graph describing retrospective provenance information. Similar to the \texttt{FairWorkflow} and \texttt{FairStep} objects, a \texttt{WorkflowRetroProv} object provides an API for accessing information stored in its RDF as well as for publishing the retrospective provenance as nanopublications (an example nanopublication of the retrospective provenance of a FAIR workflow execution is given in the footnote\footnote{\url{http://purl.org/np/RAr9-lTqsMnQrBXsmwME12mMm0AEwgcMtd9b60FOX1B_M}}).

An existing Python-based workflow automation library, \texttt{noodles}, was adopted as the core execution engine~\cite{noodles2019}. Namely, when a function is decorated with the \texttt{is\_fairstep} decorator, a \texttt{noodles Promise} is defined. Subsequently, when combining different step functions in a workflow function, these \texttt{noodles Promises} are used to construct a callgraph.

\subsection{FAIR Workflows Jupyter-Lab Extension}
The \texttt{FAIRWorkflowsExtension}~\cite{fairworkflowsextension} is a Jupyter-Lab extension to aid in the use of the \texttt{fairworkflows} library in notebooks. Upon installation it is shown as a small panel within the Jupyter Lab environment. It provides a simple GUI that allows searching of existing (nano-)published steps and workflows based on free text. These steps or workflows can then be directly injected into a cell of the notebook. In addition, you can quickly publish a step or workflow in the highlighted cell with a single button-click.

\begin{figure}
  \includegraphics[width=\linewidth]{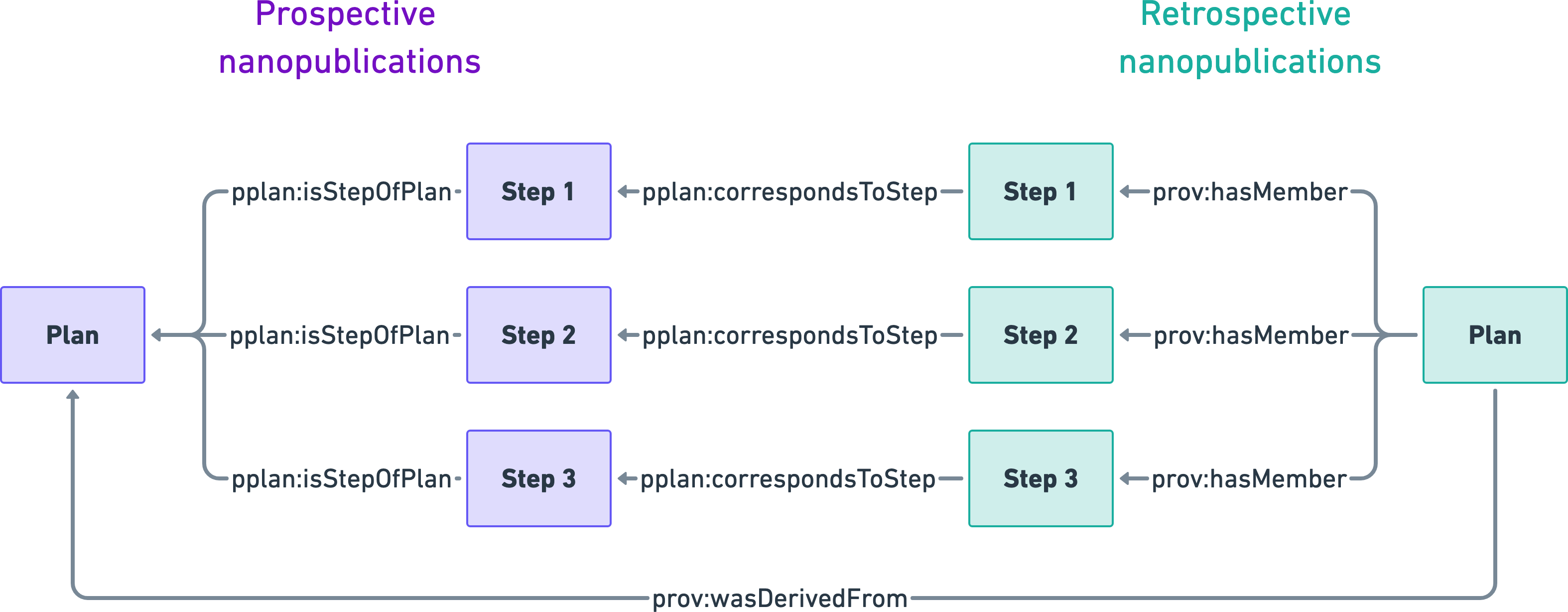}
  \caption{Prospective and Retrospective nanopublications produced by the \texttt{fairworkflows} library, based on the Plex ontology described in Ref~\cite{celebi2020towards}.}
  \label{fig:nanopubs_prospective_retrospective}
\end{figure}
\subsection{Prospective/retrospective nanopublications}
Upon publication of a workflow, a set of prospective nanopublications is produced based on the Python code that conveys the workflow; a nanopublication for each individual step and a nanopublication for the workflow as a collection of those steps. After execution of the workflow, a set of retrospective nanopublications can be published encompassing the provenance and results of running the workflow. The retrospective nanopublications are linked to the prospective nanopublications that they are derived from, step to step and workflow to workflow, as shown in Figure \ref{fig:nanopubs_prospective_retrospective}. Both retrospective and prospective nanopublications are based on the Plex ontology described in ~\cite{celebi2020towards}.

\section{User Study Design}\label{sec:userstudydesign}
A user evaluation was constructed to assess the usability and perceived FAIRness of the approach presented in the previous chapter, by students and researchers in data science related fields. Potential participants for the user evaluation were selected based on a minimum (stated) level of expertise with Python and IPython/Jupyter notebooks. The participants were approached via internal mailing lists and personal messages. A Jupyter notebook was prepared containing an introduction to the \texttt{fairworkflows} library and notebook extension, a tutorial, and three tasks involving image processing. The notebooks were containerized using Docker, with the same library dependency versions and Jupyter Lab configuration, and multiple instances were hosted remotely at Maastricht. Each container was assigned a URL and a password, each of which were communicated to different participants. The notebook file being edited by each participant was recovered as each participant completed their tasks. The evaluation notebook and all associated image files can be found in reference \cite{evaluation-notebooks-zenodo}.

\subsection{Description of user tasks}
\begin{figure}
  \includegraphics[width=\linewidth]{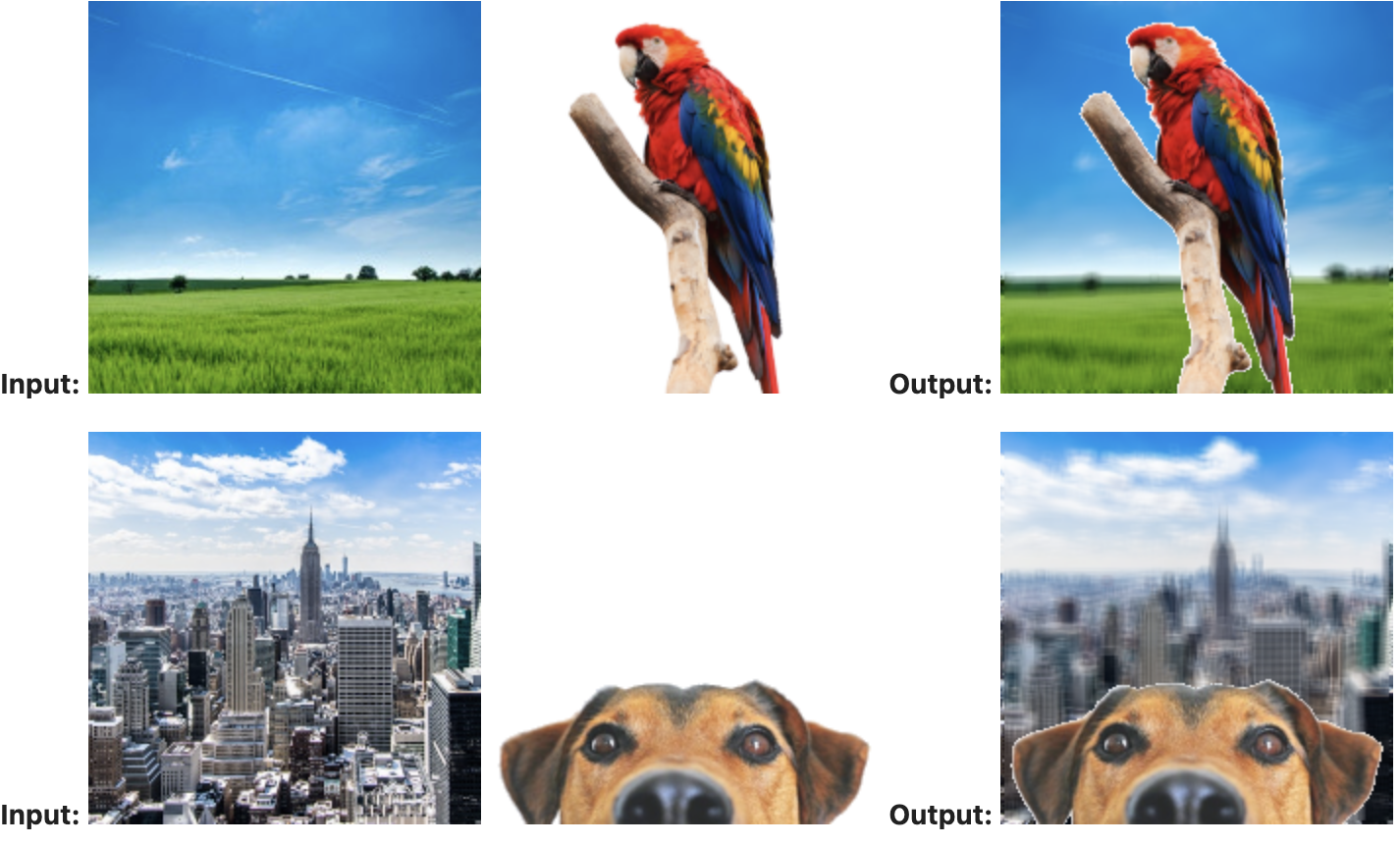}
  \caption{Task 2 of the evaluation.}
  \label{fig:fairworkflows_task2}
\end{figure}

Users were assigned three (sets of) tasks around the theme of image manipulation, such as changing the colors or adding distortions. This theme was chosen because it allows for complex multi-stage workflows, while at the same time being easy to explain and demonstrate. Task 1 and 2 involved searching for and loading previously nanopublished steps and constructing a new workflow therefrom. Task 1 required the production of a workflow that takes an input image and outputs an image that looks like a pencil sketch. The series of steps required were described in detail.
Task 2 required creation of a workflow that takes two input images and produces a combined output image, with the `foreground' image (e.g. a parrot) superimposed over a blurred version of the `background' image (e.g. a field), as illustrated in Figure \ref{fig:fairworkflows_task2}. For this task, the participants were shown these examples of input and output, but were otherwise free to choose the workflow structure. Task 3 required the participant to create their own processing step, nanopublish it, and use it in a workflow. The behaviour of the processing step and workflow in Task 3 was intentionally left as a free choice of the participant. The aim of this task was to assess whether users were able and confident enough to create new steps too.


\subsection{Questionnaire}
At the end of the study the users were directed to a questionnaire. This consisted of three questions to assess the participants' technical background, asking about prior experience with Python, Jupyter/IPython notebooks, and with computational workflow languages and management systems such as Galaxy, Knime, Taverna, or CWL. This was followed with questions about the usability of the system, following the ten standard questions of the System Usability Scale (SUS). Users were then asked four questions designed to assess their perception of the effect on FAIRness of the approach being studied. Finally, users were asked about how the approach compares with other workflow approaches, and free text boxes for general opinions on the approach. The full set of questions and response format is provided in reference \cite{evaluation-questionnaire-zenodo}.

\section{Results}\label{sec:results}

\subsection{Task Completion}
\begin{table}[tb]
\caption{Success of the participants of the user study}
\begin{tabular}{l|rrrr}
 & \multicolumn{4}{l}{number of participants who ...} \\
 & & & & partially \\
 & participated & attempted & succeeded & succeeded \\
\hline
Task 1 & 15 & 14 & 10 & 3 \\
Task 2 & 15 & 11 & 5 & 5 \\
Task 3 & 15 & 11 & 11 & 0 \\
\end{tabular}
\label{tab:task_completion}
\end{table}
Data regarding the numbers of participants attempting and successfully completing the evaluation tasks is provided in Table \ref{tab:task_completion}. In Task 1, 10 participants out of 15 succeeded in producing the canonically correct workflow, given the instructions, with a further 3 producing workflows which partially solved the problem. Of the 11 participants who attempted Task 2, 5 clearly succeeded (producing an opaque image of a tulip over a blurred mountain background) and a further 5 produced workflows with output similar to the target, but with a translucent/blended foreground, or unblurred background. In Task 3, 11 participants successfully created their own step and nanopublished it, with a variety of different implementations produced. 1 of the 15 participants did not proceed past the tutorial stage and is therefore included in the `participated' total but did not attempt any of the 3 tasks.

\subsection{Descriptive analysis of produced nanopubs}
Table~\ref{tab:fairworkflows_statistics_nps} shows the total and average number of nanopublications produced for each task. The number of nanopublications generated for each workflow's prospective and retrospective definitions is equal to the number of steps plus 1 (+1 for its plan definition). While a prospective workflow defines how the workflow should be executed, a retrospective workflow refers to how the workflow was executed with given parameters according to its definition. For Task 1, users are supposed to define 5 steps in their workflow, therefore a total of 6 nanopublications are expected to be created. However, an average of 5.5 nanopublications was published, which indicates that some steps were missing in Task 1 workflows. This is also seen for Task 2 and Task 3. For the typical 3-step workflow that solves Task 2, a total of 4 nanopublications is expected, but an average of 3.1 nanopublications was produced. This difference suggests that more people made mistakes for Task 2, which is consistent with Table \ref{tab:task_completion}.

\begin{table}[]
  \caption{Statistics of the generated nanopublications (nps)}
\begin{tabular}{ll|rrr}
                        &               & \begin{tabular}[c]{@{}r@{}}\# of users \\ who published\end{tabular} & \begin{tabular}[c]{@{}r@{}}\# total \\ nps\end{tabular} & \begin{tabular}[c]{@{}r@{}}\# average \\ nps\end{tabular} \\
\hline
\multirow{2}{*}{Task 1} & Prospective   & 13                                                                   & 71                                                                       & 5.5                                                                    \\
                        & Retrospective & 8                                                                    & 44                                                                       & 5.5                                                                    \\
\hline
\multirow{2}{*}{Task 2} & Prospective   & 11                                                                   & 35                                                                       & 3.2                                                                    \\
                        & Retrospective & 9                                                                    & 28                                                                       & 3.1                                                                    \\
\hline
\multirow{2}{*}{Task 3} & Prospective   & 11                                                                   & 30                                                                       & 2.7                                                                    \\
                        & Retrospective & 8                                                                    & 21                                                                       & 2.6                                                                   
\end{tabular}
  \label{tab:fairworkflows_statistics_nps}
\end{table}

To analyze the RDF data generated from the user study and at the same time showcasing their semantically interconnected nature, we ran a set of SPARQL queries and present their first 3 results in Table~\ref{tab:results_sparql_queroies}. The first query shown in the Listing~\ref{lst:sparql1}  returns the workflow steps and the number of times that they were used/reused in workflows. We have used the \texttt{prov:wasDerivedFrom} property to capture the information whether a step is derived from another nanopublication. With this link, we can keep track of the step lineage. 

\begin{lstlisting}[label=lst:sparql1, breaklines=true, basicstyle=\footnotesize\tt, caption={SPARQL query to list the workflow steps and how often they were used/reused in workflows},captionpos=b]
PREFIX p-plan: <http://purl.org/net/p-plan#> 
PREFIX prov: <http://www.w3.org/ns/prov#>
PREFIX rdf: <http://www.w3.org/1999/02/22-rdf-syntax-ns#>
PREFIX rdfs: <http://www.w3.org/2000/01/rdf-schema#>
PREFIX npx: <http://purl.org/nanopub/x/> 
PREFIX np: <http://www.nanopub.org/nschema#> 

SELECT ?step_label (COUNT(?plan) AS ?cnt) WHERE {
    ?step p-plan:isStepOfPlan ?plan .
    ?nanopub npx:introduces ?step .
    ?nanopub np:hasAssertion ?assertion .
    ?assertion prov:wasDerivedFrom ?step_org .
    ?step_org rdfs:label ?step_label .
} GROUP BY ?step_label ORDER BY DESC(?cnt)
\end{lstlisting}

The second query demonstrated here is listing the number of times workflow steps were executed shown in Listing~\ref{lst:sparql2}. Linking \texttt{p-plan:Activity} and \texttt{p-plan:Step} in our semantic representation through the property \texttt{p-plan:correspondsToStep} allows us to retrieve the results of this query.   

\begin{lstlisting}[label=lst:sparql2, breaklines=true, basicstyle=\footnotesize\tt, caption={SPARQL query to return how many times workflow steps were executed},captionpos=b]
PREFIX p-plan: <http://purl.org/net/p-plan#> 
PREFIX rdfs: <http://www.w3.org/2000/01/rdf-schema#>

SELECT ?step_label (COUNT(?step_prov) AS ?cnt) WHERE { 
    ?step_prov p-plan:correspondsToStep ?step .
    ?step rdfs:label ?step_label .
} GROUP BY ?step_label ORDER BY DESC(?cnt)
\end{lstlisting}

The third and final query defined in Listing~\ref{lst:sparql3} can be run to see the complexity of workflows in terms of how many unique steps are defined in a workflow.

\begin{lstlisting}[label=lst:sparql3, breaklines=true, basicstyle=\footnotesize\tt, , caption={SPARQL query to return the published workflows and how many steps they consist of},captionpos=b]
PREFIX p-plan: <http://purl.org/net/p-plan#> 
PREFIX rdfs: <http://www.w3.org/2000/01/rdf-schema#>

SELECT ?plan_label (COUNT(DISTINCT ?step) AS ?cnt)  WHERE { 
    ?step p-plan:isStepOfPlan ?plan . 
    ?plan rdfs:label ?plan_label .
} GROUP BY ?plan_label ORDER BY DESC(?cnt)
\end{lstlisting}

\begin{table}[]
  \caption{The first 3 rows of the resulting tables of the SPARQL queries defined in Listing~\ref{lst:sparql1}, \ref{lst:sparql2} and \ref{lst:sparql3}}
\begin{tabular}{p{7cm}l}
\hline
\multicolumn{2}{c}{\textbf{Query Results 1}}                                                                                                           \\ \hline
\textbf{step\_label}                                                                                                                    & \textbf{cnt}          \\ \hline
Add blur to image                                                                                                                       & 15           \\ \hline
Blend two images                                                                                                                        & 11           \\ \hline
contrast image by factor                                                                                                                & 11           \\ \hline
\multicolumn{2}{c}{\textbf{Query Results 2}}                                                                                                           \\ \hline
\textbf{step\_label}                                                                                                                    & \textbf{cnt} \\ \hline
An awesome step for creating composite image by blending images using a transparency mask & 2            \\ \hline
Add blur to image                                                                                                                       & 2            \\ \hline
Blend two images                                                                                                                        & 1            \\ \hline
\multicolumn{2}{c}{\textbf{Query Results 3}}                                                                                                           \\ \hline
\textbf{plan\_label}                                                                                                                    & \textbf{cnt}          \\ \hline
takes an images and returns a pencil sketch version  & 5            \\ \hline
Convert an image to a pencil sketch & 5            \\ \hline
A workflow converts an image to a pencil sketch & 5            \\ \hline
\end{tabular}
 \label{tab:results_sparql_queroies}
\end{table}

\subsection{Usability}
In total there were 15 participants in the evaluation of which 14 completed the end-of-evaluation questionnaire, with the majority indicating intermediate (50\%) or expert (35.7\%) experience with Python. Similarly, 71.4\% of participants indicated intermediate or expert prior experience with Jupyter/iPython notebooks (35.7\% in each category). On the other hand, 64.3\% participants had almost no experience with other computational workflows systems (broad examples given were Galaxy, Knime, Taverna, CWL etc) with only 21.4\% claiming intermediate or expert experience.

The usability of the system was assessed with the ten standard System Usability Scale (SUS) questions. The user responses to these questions are summarized in Table \ref{tab:sus-result}.
\begin{table*}[tb]
\centering
\caption{SUS scores derived from participant responses to standard questions.}
\label{tab:sus-result}
\small
\begin{tabular}{lr|r@{~~~~}r@{~~~~}r@{~~~~}r@{~~~~}r|r}
 & & \multicolumn{5}{c|}{better $\rightarrow$} & \\
 & odd questions: & \phantom{00}1 & \phantom{00}2 & \phantom{00}3 & \phantom{00}4 & \phantom{00}5 &  \\
SUS questions: & even questions: & 5 & 4 & 3 & 2 & 1 & score \\
\hline
\multicolumn{2}{l|}{\small{1: I think that I would like to use this system frequently.}} & 0 & 0 & 3 & 7 & 4 & 76.79 \\
\multicolumn{2}{l|}{\small{2: I found the system unnecessarily complex.}} & 0 & 1 & 2 & 4 & 7 & 80.36 \\
\multicolumn{2}{l|}{\small{3: I thought the system was easy to use.}} & 0 & 0 & 2 & 7 & 5 & 80.36 \\
\multicolumn{2}{l|}{\small{4: I think that I would need the support of a technical person to be able to use this system.}} & 1 & 1 & 0 & 3 & 9 & 82.14 \\
\multicolumn{2}{l|}{\small{5: I found the various functions in this system were well integrated.}} & 0 & 1 & 2 & 5 & 6 & 78.57 \\
\multicolumn{2}{l|}{\small{6: I thought there was too much inconsistency in this system.}} & 0 & 0 & 4 & 3 & 7 & 80.36 \\
\multicolumn{2}{l|}{\small{7: I would imagine that most people would learn to use this system very quickly.}} & 0 & 0 & 6 & 5 & 3 & 69.64 \\
\multicolumn{2}{l|}{\small{8: I found the system very cumbersome to use.}} & 0 & 0 & 3 & 5 & 6 & 80.36 \\
\multicolumn{2}{l|}{\small{9: I felt very confident using the system.}} & 0 & 0 & 3 & 6 & 5 & 78.57 \\
\multicolumn{2}{l|}{\small{10: I needed to learn a lot of things before I could get going with this system.}} & 1 & 1 & 0 & 4 & 8 & 80.36 \\
\hline
 & total: & 2 & 4 & 25 & 49 & 60 & 78.75 \\
\end{tabular}
\end{table*}
Our system achieved an overall SUS score of 78.75, which is above the average score reported in the literature (70.14) and is somewhere in the middle between ``good'' and ``excellent'' on an adjective scale \citep{bangor2008empirical}. These results indicate that our system is easy to use, and that users feel confident and satisfied using our system.

\subsection{Perceived effect on FAIR}
\begin{table}[tb]
\caption{The perceived effect of the approach on FAIRness, with regards to workflow steps, as judged by participants answering the questionnaire.}
\begin{tabular}{l|lllll|l}
Perceived effect on... & 1 & 2 & 3 & 4 & 5 & average \\
\hline
Findability & 0 & 0 & 1 & 5 & 8 & 4.5\\
Accessibility & 0 & 0 & 3 & 6 & 5 & 4.1 \\
Interoperability &0 & 0 & 5 & 4 & 5 & 4.0 \\
Reusability & 0 & 0 & 0 & 4 & 10 & 4.7 \\
\end{tabular}
\label{tab:perceived_fairness}
\end{table}

The perceived effect of this approach on FAIRness was also assessed by means of 4 questions asking how useful the participant thought the system was for making workflow steps more findable, accessible, interoperable and reusable. The results, graded from 1 to 5, are presented in Table \ref{tab:perceived_fairness}. The average responses in all four categories were above 4, with findability and reusability scoring most highly.

\subsection{Qualitative analysis of participants' responses}
In response to a question about what they \textit{liked} most about the system, respondents indicated the ease of publishing their own steps and finding those of others using the provided search interface that integrated smoothly with the Jupyter environment. The ability to write standard python code and make it fair with simple application of a python decorator was also appreciated. Regarding what was most \textit{disliked}, respondents indicated the very prototypical state of the toolset, the requirement for good knowledge of python and workflows, and the lack of verification of published code i.e. broken code could be nanopublished, which is frustrating for users searching for other steps, or in accidentally publishing their own code multiple times.

\section{Discussion}\label{sec:discussion}

We presented a set of tools that together allow for composing, reusing and executing semantically annotated workflows, with the aim of minimal adjustment of a regular data science workflow in a Jupyter Notebook environment. The user study results demonstrate that the developed tools are sufficient to achieve multi-stage workflow creation and publication, while attaining FAIR principles. Most users reported feeling confident and satisfied using our system, and usability was graded between ``good'' and ``excellent''.

\subsection{Availability of software}
All software within our toolset (the \texttt{nanopub} and \texttt{fairworkflows} Python libraries, and the \texttt{FAIRWorkflowsExtension} plugin for Jupyter Lab) is available on GitHub\footnote{\url{https://github.com/fair-workflows}}, easily installable via the Python Package Index. It is released as free open-source software, so it can be extended by the community, hopefully inciting other promising endeavors for the creation of FAIR workflows.

\subsection{Current limitations of step reusability}
The \texttt{fairworkflows} library publishes the raw python code as part of the produced nanopublications, which is an evident security risk if it is fetched and run uncritically on another machine. While the environment (python version and required dependencies) can be included in the nanopublication, the library does not currently enforce it, therefore it does not guarantee that a published step can be reused in a different environment (a key limitation in the FAIR context). This problem can potentially be resolved in future work through e.g. containerization of a computational step. However, there is a balance to be struck with regards to the attractive simplicity of a Jupyter Notebook. The Python code serves as a detailed description of a computational step, and the injection of raw Python code in Jupyter notebooks that is then editable provides a powerful way to mix and match steps in a Jupyter notebook without disrupting the user experience.

\subsection{Limitations of user study}
While the majority of participants had intermediate or expert experience with Python and Jupyter notebooks prior to the study, relatively few ($\sim$20\%) had experience with formal computational workflow systems and languages. Thus the responses to questions regarding comparison with other workflow systems are not covered in our analysis here.

As the participants of this experiment worked on tasks that were given by us, it remains to be seen to what extent our results are still valid when they use the system for their own workflows and in their own environments. We could not find any indication in the presented results that our approach would not work in such a real setting, but this needs additional future studies to be confirmed.

\subsection{Future work}
The \texttt{fairworkflows} library has some features that were not assessed in this particular evaluation but which can throw light on the general approach envisaged here. Semantic types may be set for all input and output variables of any steps, via arguments (of the same name) to the decorator. These are then used in any generated RDF, and in principle could be used in future for type comparison operations. There is also a prototype ``Manual Task Assistant''. The \texttt{is\_fairstep} decorator can be used to mark a function as a manual ``step'' by setting \texttt{is\_manual\_task} to \texttt{True}. Manual step execution triggers the opening of a web interface that guides the user through the manual task, thereby collecting input from the user that can be used in subsequent steps of the workflow.

The notebook extension developed here is a proof-of-concept, developed partly on the basis that more initiatives are needed to make production of FAIR scientific workflows more intuitive through the provision of user-friendly interfaces to RDF.

\section{Conclusion}
We have presented an approach to applying FAIR principles to scientific workflows in Jupyter notebooks, using semantic workflow descriptions and automatic generation of prospective and retrospective provenance as nanopublications. The results of our evaluation of this proof-of-concept, involving 15 participants, shows that our approach is feasible and that such a system can be made user-friendly and efficient to use. This work is intended to be extended further with semantic typing and manual task support (already present in the \texttt{fairworkflows} library and supported by the Plex ontology) and to resolve some of the techincal interoperability and security concerns through containerisation.

In general, our approach could pave the way for a global FAIR ecosystem of workflows and workflow steps, where such computational components can be flexibly and reliably discovered, connected, executed, and traced back. This in turn would allow us to strongly connect input datasets of computational processing with their output datasets, thereby providing detailed and semantically precise links between these datasets. These in turn can then be used to further improve dataset discovery and analysis. We therefore believe that these technologies form an important contribution to put data-focused research on a more solid and more sustainable foundation.

\begin{acks}
This work was supported by the Dutch Research Council (NWO) (No. 628.011.011) and the Netherlands eScience Center (No. NLeSC P 17.0201).
We would like to thank Vincent Emonet for his invaluable help in setting up and hosting the evaluation notebook containers.
\end{acks}

\bibliographystyle{ACM-Reference-Format}
\bibliography{paper}

\end{document}